\documentclass[%
superscriptaddress,
amsmath,amssymb,
 aps,prd,
 twocolumn,
 tightenlines,
]{revtex4-1}

\usepackage{graphicx,url,amssymb,amsmath,longtable,rotating,color,units,wasysym,subfigure,epsfig,epstopdf}
\usepackage[dvipsnames]{xcolor}
\usepackage{enumitem}

\usepackage{graphicx}
\usepackage{dcolumn}
\usepackage{bm}
\usepackage{url, hyperref}
\usepackage{array,multirow}
\usepackage{color} 
\usepackage{aas_macros} 
\usepackage{ifthen}
\usepackage{etoolbox}

\usepackage{siunitx} 
\usepackage{acronym,cleveref}
\usepackage[dvipsnames]{xcolor}
\usepackage[normalem]{ulem} 

\newtoggle{fullauthorlist}
\toggletrue{fullauthorlist}

\newtoggle{endauthorlist}
\toggletrue{endauthorlist}

\definecolor{darkgreen}{rgb}{0.0, 0.5, 0.0}

\newif\ifcomments
\commentstrue

\newcommand{\Ogw}{\Omega_\text{gw}}

\newcommand{\lmax}{l_\text{max}}

\newcommand{\fref}{f_\text{ref}}

\newcommand{\rundates}{{16:00:00 UTC on 30 November, 2016 to 22:00:00 UTC on 25 August, 2017}}

\newcommand{\cutDays}{{16\%}~}
\newcommand{\livetime}{{$\unit[99]{\text{days }}$}}

\acrodef{LIGO}{Laser Interferometer Gravitational-wave Observatory}
\acrodef{aLIGO}{Advanced LIGO}
\acrodef{GW}{gravitational waves}
\acrodef{CGB}{cosmological gravitational-wave background}
\acrodef{SNR}{signal-to-noise ratio}

\begin{document}

\title{Directional limits on persistent gravitational waves \\
using data from Advanced LIGO's first two observing runs}
\date{\today}

\noaffiliation

\iftoggle{endauthorlist}{
  %
  %
  \let\mymaketitle\maketitle
  \let\myauthor\author
  \let\myaffiliation\affiliation
  \author{The LIGO Scientific Collaboration}
  \author{The Virgo Collaboration}
\noaffiliation
}{
  %
  %
  \iftoggle{fullauthorlist}{
  }{
    \author{The LIGO Scientific Collaboration}
    \affiliation{LSC}
    \author{The Virgo Collaboration}
    \affiliation{Virgo}
  }
}

\begin{abstract}
We perform an unmodeled search for persistent, directional gravitational wave (GW) sources using data from the first and second observing runs of Advanced LIGO.  
We do not find evidence for any GW signals.  
We place limits on the broadband GW flux emitted at 25~Hz from point sources with a power law spectrum at 
$F_{\alpha,\Theta} <(0.05-25)\times 10^{-8} ~\unit[]{erg\,cm^{-2}\,s^{-1}\,Hz^{-1}}$ and the (normalized) energy density spectrum in GWs at 25 Hz from 
extended sources at
$\Omega_{\alpha}(\Theta) <(0.19-2.89)\times 10^{-8} ~\unit[]{sr^{-1}}$ where $\alpha$ is the spectral index of the energy density spectrum.
These represent improvements of $2.5-3\times$ over previous limits.
We also consider point sources emitting GWs at a single frequency, 
targeting the directions of Sco X-1, SN 1987A, and the Galactic Center. The best upper limits on the strain amplitude of a potential source in these three directions range from $h_0 < (3.6-4.7)\times 10^{-25}$, 1.5$\times$ better than previous limits set with the same analysis method.
We also report on a marginally significant outlier at 36.06~Hz.
This outlier is not consistent with a persistent gravitational-wave source as its 
significance diminishes when combining all of the available data.
\end{abstract}

\maketitle


\acrodef{BNS}{Binary Neutron Star}
\acrodef{BBH}{Binary Black Hole}
\acrodef{SNR}{signal-to-noise ratio}
\acrodef{GW}{gravitational wave}

\section{Introduction}\label{sec:intro}
The stochastic gravitational wave (GW)
background (SGWB) is the superposition of many sources of GWs
in the Universe~\cite{Christensen:2018iqi}. Anisotropies in the SGWB can be generated by spatially extended sources such as a population of neutron
stars in the galactic plane or a nearby galaxy
\cite{Talukder:2014eba,Mazumder:2014fja}, or from perturbations in
statistically-isotropic backgrounds formed at cosmological distances such as the
compact binary background
\cite{Jenkins:2018uac,jenkins_agwb_bh_distributions,Cusin:2018rsq,Cusin:2017mjm,Cusin:2017fwz,contaldi_anisotropy_line_of_site}
or the background from cosmic strings \cite{Jenkins:2018lvb}. Cross-correlation
based methods have been used to search for the anisotropic background in
previous observing runs \cite{o1_directional,sph_results,Renzini_mapping_methods,Renzini_o1_mapping_results} of the initial and
Advanced Laser Interferomter Gravitational-wave Observatory
(LIGO)~\cite{AdvancedLIGO}, and future searches will incorporate data from the
Advanced Virgo~\cite{AdvancedVirgo} detector. Using very similar techniques,
one can also search for point sources with an unknown phase evolution, which
could include rotating neutron stars in the Galaxy
\cite{radio_method,radiometer}. Since a SGWB search is by nature
un-modelled, performing the anisotropic SGWB search allows us to take an
eyes-wide-open approach to exploring the GW sky.

In this paper, we present the results of three complementary searches, which
probe different types of anisotropy. All of the searches are based on
cross-correlation methods; for a review see \cite{RomanoCornish}. A spherical
harmonic decomposition (SHD) of the GW power on the
sky~\cite{sph_methods,sph_results} is optimized to search for extended sources
on the sky with a smooth frequency spectrum. The broadband radiometer
analysis~\cite{radio_method,radiometer} (BBR) is optimized for detecting
resolvable, persistent point-sources emitting GWs across a wide frequency band.
Finally, the directed narrowband radiometer (NBR) looks
at the frequency spectrum for three astrophysically
interesting directions: Scorpius X-1 (Sco
X-1)~\cite{scox1_cross_corr_o1_results,scox1_viterbi_o1_results}, Supernova
1987A (SN 1987A)~\cite{SunEA_1987a,ChungEA_sn1987a}, and the Galactic
Center~\cite{Dergachev_o1_loose_gc, gc_2013}.
We do not find a significant detection for any of the
searches, and so we place upper limits on the amplitude of the anisotropic
SGWB, and on point sources with broad and narrow frequency
ranges. Our upper limits improve on the best results from previous runs
\cite{o1_directional} by approximately a factor of 2.5-3 for the broadband
searches and a factor of 1.5 for the narrowband searches.
For the narrowband radiometer search, we find a marginally
significant outlier in the direction of SN 1987A,
when analyzing just the data from LIGO's second observing run (O2).
Its significance diminishes, however, when including all of the available data.


\section{Data}\label{sec:data}
We analyze strain data from the first (O1) and
second (O2) observing runs of Advanced LIGO's $\unit[4]{km}$ detectors in
Hanford, Washington (H1) and Livingston, Louisiana (L1). The O1 data set used here was collected from 15:00 UTC on 18
September, 2015 to 16:00 UTC on 12 January, 2016, while the O2 data set was collected from
\rundates. In O2, linearly coupled noise was removed
from the strain time series at H1 and L1 using Wiener filtering
\cite{Noise_sub_O2,Wiener_O2,Noise_sub_1,Noise_sub_2,Noise_sub_3}.
The Virgo (V1) detector started to collect data from August 2017 but
does not contribute significantly to the sensitivity of SGWB searches in
O2, both because its noise level is much higher than the LIGO detectors and
because it ran for a much shorter period of time. Therefore, we do not include
Virgo in this analysis. We plan, however, to include Virgo in the analysis of
data from future observation runs.  

Our data processing methods follow the procedure used in O1
\cite{o1_directional,o1iso}. First, we down-sample the strain time series from
\unit[16,384]{Hz} to \unit[4,096]{Hz}. We then divide the data into
$\unit[192]{s}$, 50\% overlapping, Hann-windowed segments, and apply a
cascading 16th order Butterworth digital high-pass filter with a knee frequency of
$\unit[11]{Hz}$. We compute the cross correlation of coincident $\unit[192]{s}$
segments at both detectors in the frequency domain, and then coarse-grain to a
frequency resolution of $\unit[1/32]{Hz}$. Finally, we optimally combine
results from those overlapping time segments to produce the final
cross-correlation estimate~\cite{allen-romano}.

In order to account for non-Gaussian features in the data, we remove segments
associated with instrumental artifacts and hardware injections used for signal
validation~\cite{gw150914_detchar, o1hwinj}.  Segments containing known
GW signals \cite{gwtc_1} are also excluded.  Finally, we apply
a non-stationarity cut (see, e.g.,~\cite{stoch-S5}) to eliminate segments where
the power spectral density of the noise changes on time scales that are of the
same order as the chosen segment length.  In total these cuts removed \cutDays
of the data, leading to a total search live-time of~\livetime from the O2 run.
For our results where we combine data between the O1 and O2 observing runs we
have a total search livetime of 129 days.  In addition, frequency bins
associated with known instrumental artifacts are
removed~\cite{o1_o2_lines_paper}. These frequency domain cuts discarded~4\% of
the most sensitive frequency band for the BBR and SHD searches and 15\% of the
observing band for the NBR search. The subtraction of linearly coupled noise did
not introduce any new frequency domain cuts.

 The broadband searches integrate over frequencies between 20 and 500 Hz. This
range accounts for more than $99\%$ of the sensitivity for the power law spectral models that we use (see Table 1 of~\cite{o2_isotropic_paper}).  The narrowband analysis searches over the frequency band
from 20 to 1726 Hz using frequency bins of various sizes depending upon
frequency and sky direction. The lower edge of this range is chosen because of increased noise
and non-stationarity at lower frequencies, while the upper edge of the range is a product of the
filter used to resample the data from 16,384 Hz to 4,096 Hz.

 \begin{table*}
    \begin{tabular}{ccc | cc ||cc|cc || cc }
    \multicolumn{7}{c}{}& \multicolumn{4}{c}{\bf All-sky (broadband) Results} \\
   \multicolumn{5}{c}{}& \multicolumn{2}{c}{Max SNR  (\% $p$-value)} & \multicolumn{2}{c}{Upper limit ranges } & \multicolumn{2}{c}{O1 Upper limit ranges }\\
     \hline
     &$\alpha$&  & $\Ogw$ & $H(f)$ & BBR &  SHD   & BBR ($\times 10^{-8}$) & SHD ($\times 10^{-8}$) & BBR ($\times 10^{-8}$) & SHD ($\times 10^{-8}$) \\
     \hline 
     &0&			& constant 			&  $\propto f^{-3}$        & 3.09  (9) & 2.98 (9) & 4.4 -- 25   & 0.78 -- 2.90 &  15 -- 65   & 3.2 -- 8.7  \\
     &$2/3$& 	& $\propto f^{2/3}$	&  $\propto f^{-7/3}$      & 3.09 (20) & 2.61 (31) & 2.3 -- 14  & 0.64 --2.47 & 7.9 -- 39  & 2.5 -- 6.7  \\
     &3&			& $\propto f^{3}$	&  constant       		 & 3.27 (66) & 3.57 (27) & 0.05 -- 0.33 & 0.19 -- 1.1 & 0.14 -- 1.1 & 0.5 -- 3.1  \\
     \hline
   \end{tabular}

\caption{Search information for BBR and SHD. On the left side of the table we show the value of the power-law spectral index, $\alpha$, and the scaling of $\Ogw$ and $H(f)$ with frequency. 
To the right we show results for the broadband radiometer (BBR) and spherical harmonic decomposition (SHD) searches for the combined O1 and O2 analysis, as well as the results from O1 for comparison. We show the maximum SNR across all sky positions for each spectral index, as well as an estimated $p$-value. We also show the range of 95\% upper limits on energy flux set by the BBR search across the whole sky [${\text{erg cm}^{-2}\text{ s}^{-1} \text{ Hz}^{-1}}$] and the SHD range of upper limits on normalized energy density across the whole sky [${\text{sr}^{-1}}$]. These limits use data from both O1 and O2. The median improvement across the sky compared to limits set in O1 is 2.6-2.7 for the BBR search and 2.8-3 for the SHD search, depending on power-law spectral index.}\label{tab:alphas}

 \end{table*}


\section{Methods}\label{sec:methods}
The anisotropic SGWB background can be defined in terms of the dimensionless energy density $\Omega_{\rm gw}(f,\Theta)$ per unit frequency $f$ and solid angle $\Theta$,
\begin{equation}
    \Ogw(f,\Theta)=\frac{f}{\rho_c}\frac{{\rm d}^3 \rho_{GW}}{{\rm d}f {\rm d}^2 \Theta},
\end{equation}
 where $\rho_c=3 H_0^2c^2/(8\pi G)$ is the critical energy density needed to have a spatially flat Universe. We take the Hubble constant to be $H_0=\unit[67.9]{km\,s^{-1}\,Mpc^{-1}}$~\cite{Planck_2015}.
Following past analyses, we assume that we can factorize $\Ogw$ into frequency and sky-direction dependent terms,
\begin{equation}
    \label{eq:omgw_direction_definition}
    \Ogw (f, \Theta) = \frac{2\pi^2}{3H_0^2}f^3 H(f)\mathcal{P}(\Theta).
\end{equation}
This quantity has units of the dimensionless energy density parameter per steradian. For the radiometer searches it is useful to define a different representation in terms of energy flux,
\begin{equation}
    \label{eq:flux_definition}
    \mathcal{F}(f, \Theta) = \frac{c^3\pi}{4G} f^2 H(f) \mathcal P(\Theta),
\end{equation}
which has units of $\unit[]{erg\,cm^{-2}\,s^{-1}\,Hz^{-1}\,sr^{-1}}$, where $c$ is the speed of light and $G$ is Newton's gravitational constant. 

We divide the searches into the \emph{broadband} searches (SHD and BBR), which
produce sky maps where the flux has been integrated over a broad range of
frequencies, and the $\emph{narrowband}$ search (NBR), which looks at the
strain amplitude spectrum in a fixed sky direction. For the broadband searches,
we typically assume that the energy spectrum has a power law form,
$H(f)=(f/f_{\rm ref})^{\alpha-3}$, where $\alpha=\{0,2/3,3\}$ describes a range of
astrophysical and cosmological models~\cite{o1_directional}, and $f_{\rm ref}$
is a reference frequency which we take to be 25 Hz, as
in~\cite{o1_directional}. The SHD search looks for sources with a large angular
extent. We express the results in terms of the spherical harmonic
decomposition of ${\Omega_{\rm gw}(f,\Theta)}$ assuming a power-law in frequency of spectral index $\alpha$. We then report the energy density in each direction at a reference frequency of 25~Hz, denoted by $\Omega_\alpha(\Theta)$.

For the BBR search, we assume that the
angular distribution of the power is localized in a $\unit[1]{deg^2}$ pixel,
$\mathcal{P}(\Theta)=\mathcal{P}_{\Theta_0} \delta^2(\Theta,\Theta_0)$. The
results of the BBR are then given in terms of the quantity
$F_{\alpha,\Theta_0}$, which is the flux evaluated at the reference frequency of 25 Hz, assuming a power law, after integrating over solid angle. The explicit definitions of $F_{\alpha,\Theta_0}$ and $\Omega_{\alpha}(\Theta)$ are given in the Technical Supplement.

Finally, the NBR search does not integrate over frequency, and
attempts to measure the strain amplitude, $h_0$, of a putative monochromatic
source in each frequency bin independently. This includes combining
adjacent 0.031~Hz frequency bins together to account for the Doppler modulation
due to the motion of the Earth around the solar system barycenter and any binary
motion of the source itself~\cite{o1_directional}.

The full description of the methods used to search for an anisotropic
SGWB is presented in the Technical Supplement and in the paper describing the analysis of the
Advanced LIGO O1 data. We follow the notation presented in that
Letter~\cite{o1_directional}. 

The searches all generally start by estimating the dirty map $X_\nu$, and its
corresponding covariance matrix $\Gamma_{\mu\nu}$, referred to here as the
Fisher matrix~\cite{sph_methods,o1_directional,Mitra_2008_radio_method2}. The
dirty map represents an estimate of the GW power as seen through the detector's
beam matrix. 

Given the Fisher matrix $\Gamma^{I}_{\mu\nu}$ and dirty map $X^I_\nu$, where
$I$ labels the observing run, we can form a combined Fisher matrix and dirty
map by summing the results from the two runs, O1 and O2 \cite{RomanoCornish}
\begin{eqnarray}
    \Gamma_{\mu\nu}& =& \Gamma^{\rm (O1)}_{\mu\nu} + \Gamma^{\rm (O2)}_{\mu\nu}, \nonumber \\
    X_\mu &=& X^{\rm (O1)}_{\mu}+X^{\rm (O2)}_\mu.
\end{eqnarray}
From the combined Fisher matrix and dirty map, we can construct estimators of the power on the sky via:
\begin{equation}
    \hat{\cal P}_\mu = \sum_\nu\,\left(\Gamma_{\rm R}^{-1}\right)_{\mu\nu} X_{\nu}\,.
\end{equation}
In the above equations, $\mu$, $\nu$ label either pixels (i.e., directions on
the sky) or spherical harmonic components---i.e., $\mu\equiv (lm)$, depending
on which basis is used to represent the sky maps.  The subscript `R' on the
Fisher matrix means that regularization has been applied (e.g., singular value
decomposition) in order to perform the matrix inversion~\cite{o1_directional}.

We can also construct an estimate of the angular power spectrum, $C_l$, for the
SGWB from the estimate of the spherical harmonics
coefficients, $\hat{\mathcal P}_{lm}$. The $C_l$'s describe the angular scale
of the structure found in the clean maps~\cite{sph_methods}
\begin{equation}
    \hat C_{l} = \left(\frac{2\pi^2 \fref^3}{3H_0^2}\right)^2\frac{1}{1 + 2 l} \sum_{m=-l}^{l} \left[|\hat{\mathcal P}_{lm}|^2 - (\Gamma_{\rm R}^{-1})_{lm,lm}\right].
\end{equation}

We have also used theoretical models for the SGWB from compact
binaries~\cite{Jenkins:2018uac} and from Nambu-Goto
cosmic strings~\cite{Jenkins:2018lvb} to check our assumption that the SGWB energy
density $\Omega_\textrm{gw}(f,\Theta)$ can be factorised into a spectral shape
term and an angular power term.  
We find that both models predict $C_l$'s that follow the
appropriate frequency power laws ($\alpha=2/3$ for compact
binaries and $\alpha=0$ for cosmic strings) across the frequency
range in which the LIGO stochastic searches are most sensitive,
thereby supporting this assumption 
(see also~\cite{cusin_source_dependencies}).

\section{Results}\label{sec:results}
\subsection{Broadband radiometer and spherical harmonic decomposition results}\label{ssec:results:bbr}
\begin{figure*}[htbp!]
  \begin{tabular}{ccc}
  	\includegraphics[width=2.15in]{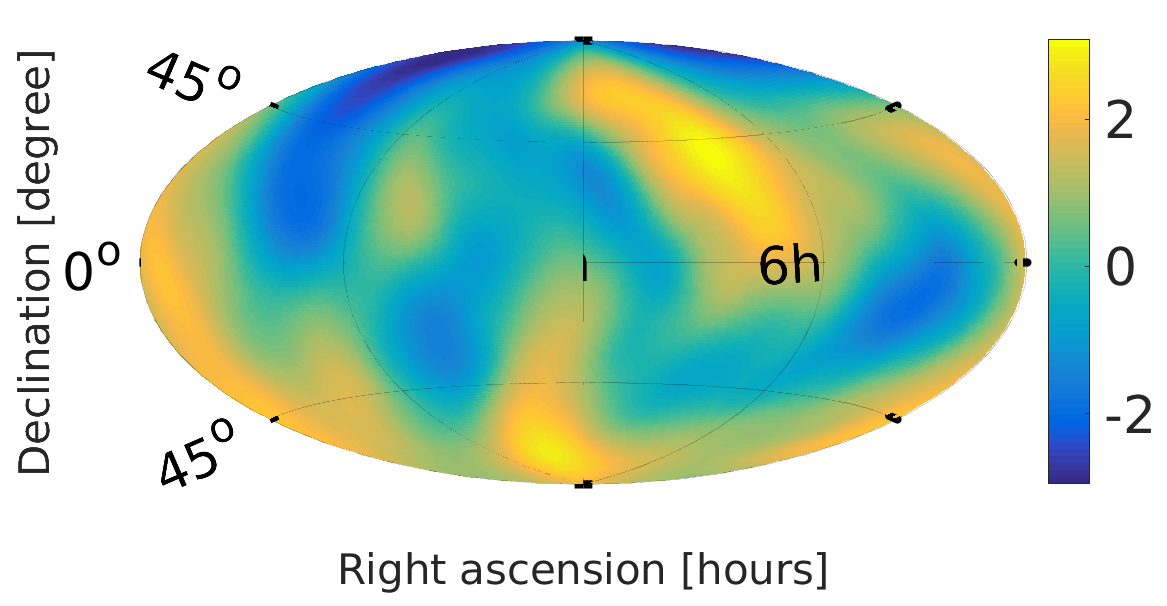} \hspace{.4cm} &
  	\includegraphics[width=2.15in]{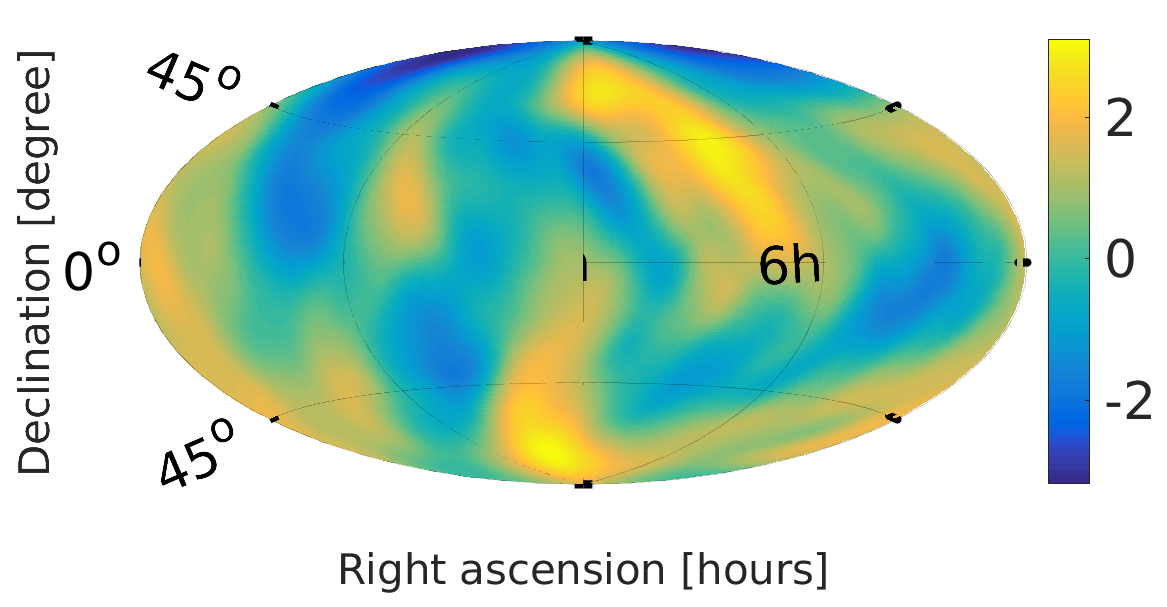} \hspace{.4cm} &
  	\includegraphics[width=2.15in]{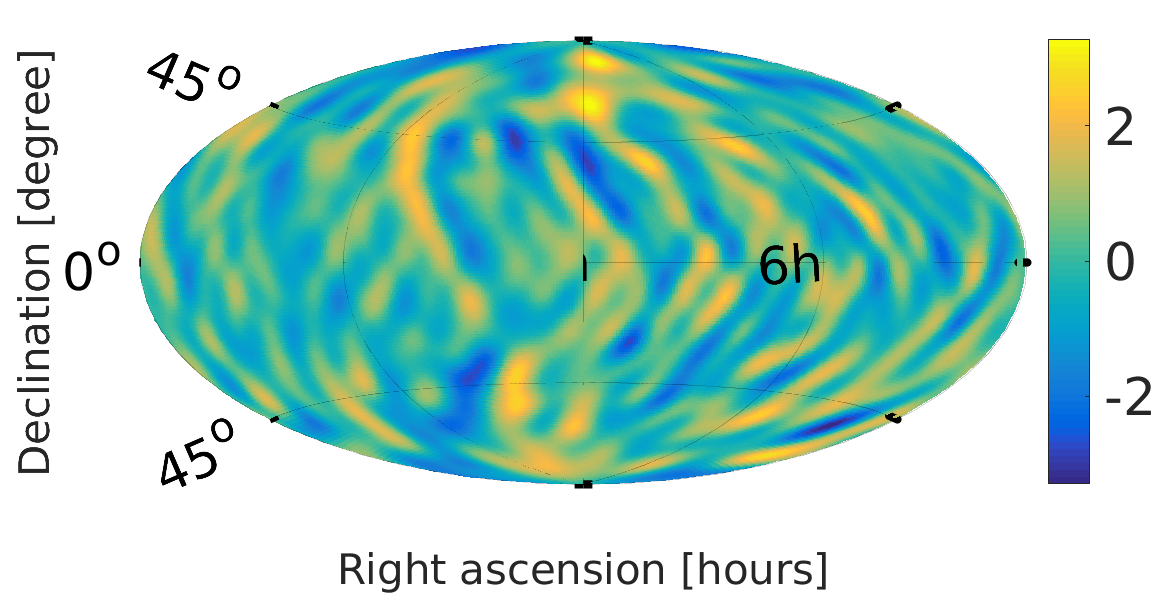} \hspace{.4cm} \\
  	\includegraphics[width=2.15in]{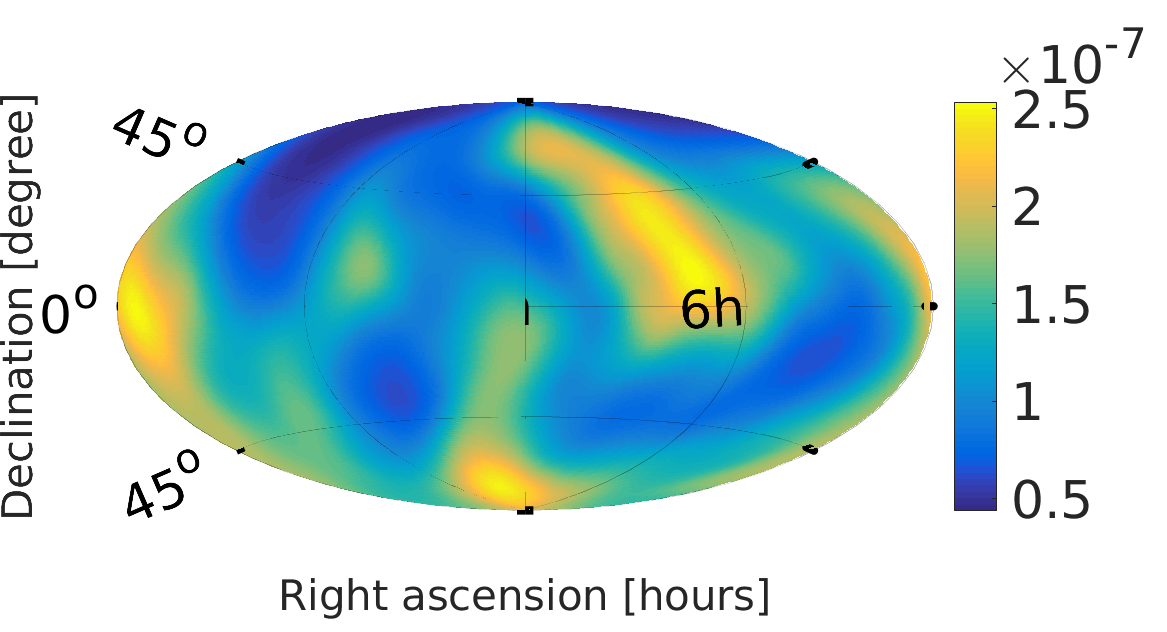} &
  	\includegraphics[width=2.15in]{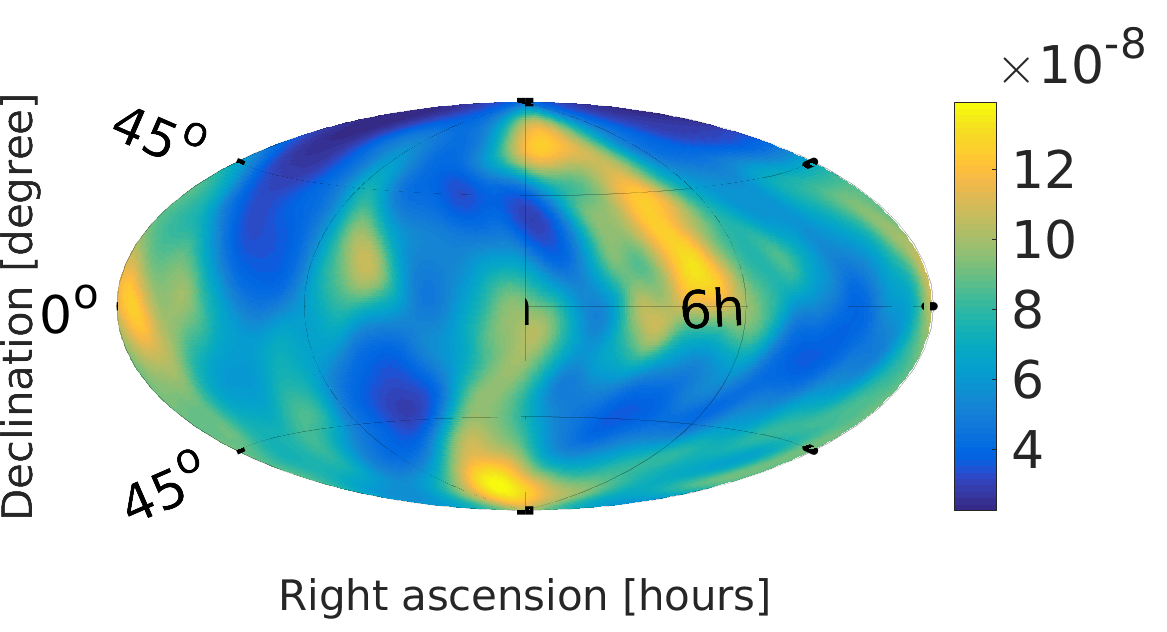} &
  	\includegraphics[width=2.15in]{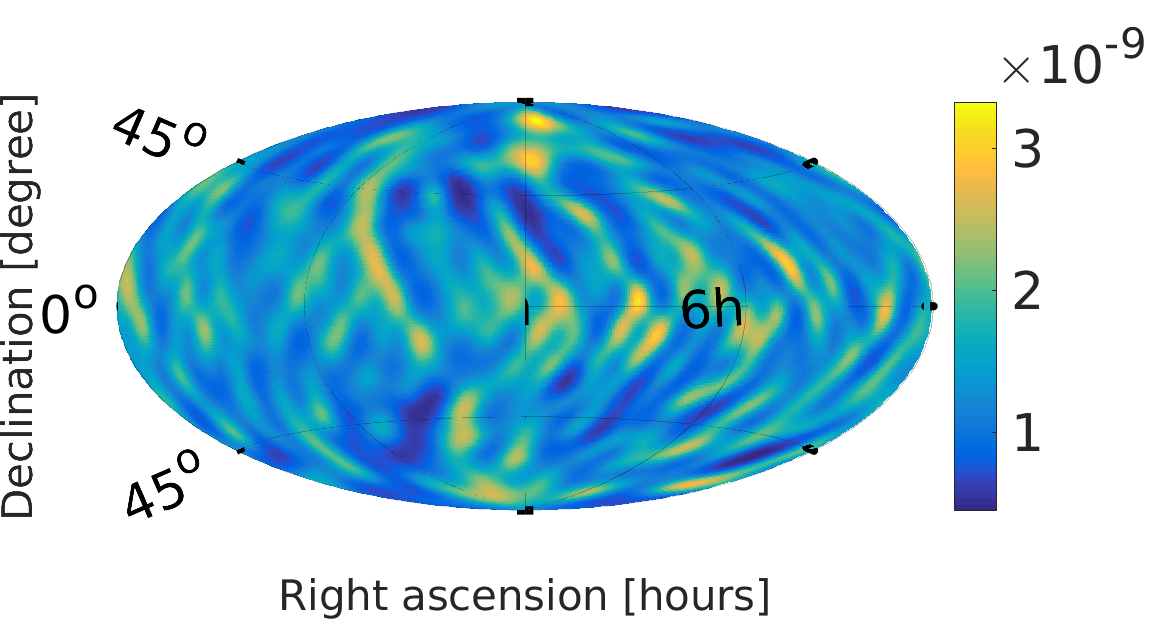} \\
  \end{tabular}
  \caption{
Broadband radiometer maps illustrating a search for point-like sources. The top
row shows maps of SNR, while the bottom row shows maps of the upper limits at
95\% confidence on energy flux $F_{\alpha,\Theta_0}$ [$\text{erg cm}^{-2}\text{
s}^{-1} \text{ Hz}^{-1}$]. Three different power-law indices, $\alpha = 0, \,
2/3$ and 3, are represented from left to right.  The $p$-values associated with
the maximum \ac{SNR} are (from left to right) $p=9\%,\,p=20\%,\,p=66 \%$ (see
Table \ref{tab:alphas}). }  

\label{fig:RADmaps}  
\end{figure*}

\begin{figure*}[htbp!]
  \begin{tabular}{ccc}
	\includegraphics[width=2.15in]{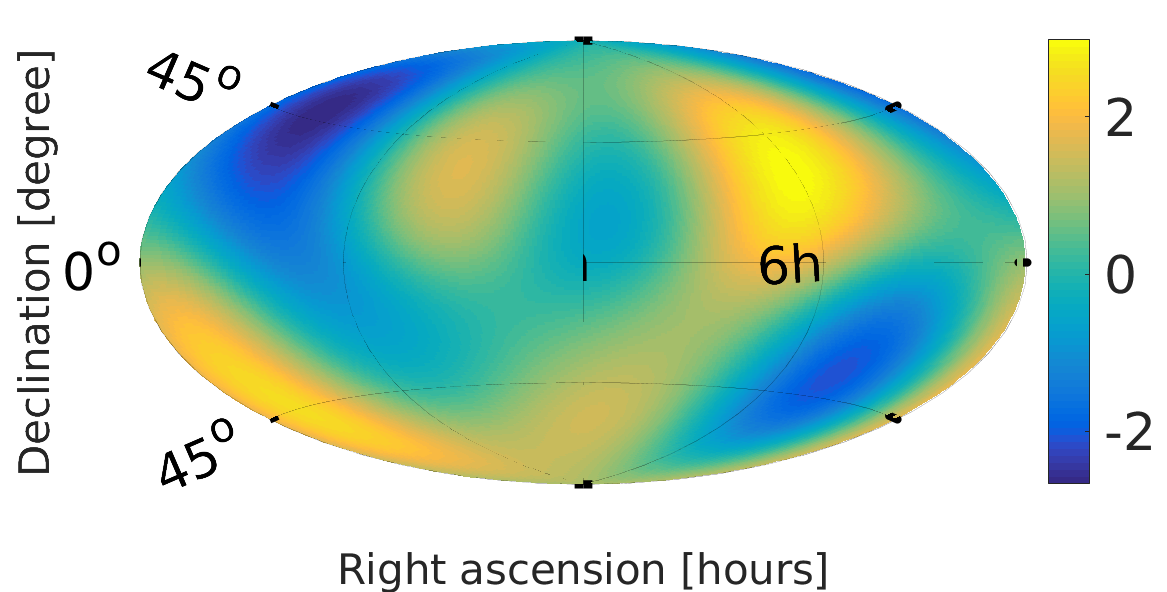} \hspace{.4cm} & 
	\includegraphics[width=2.15in]{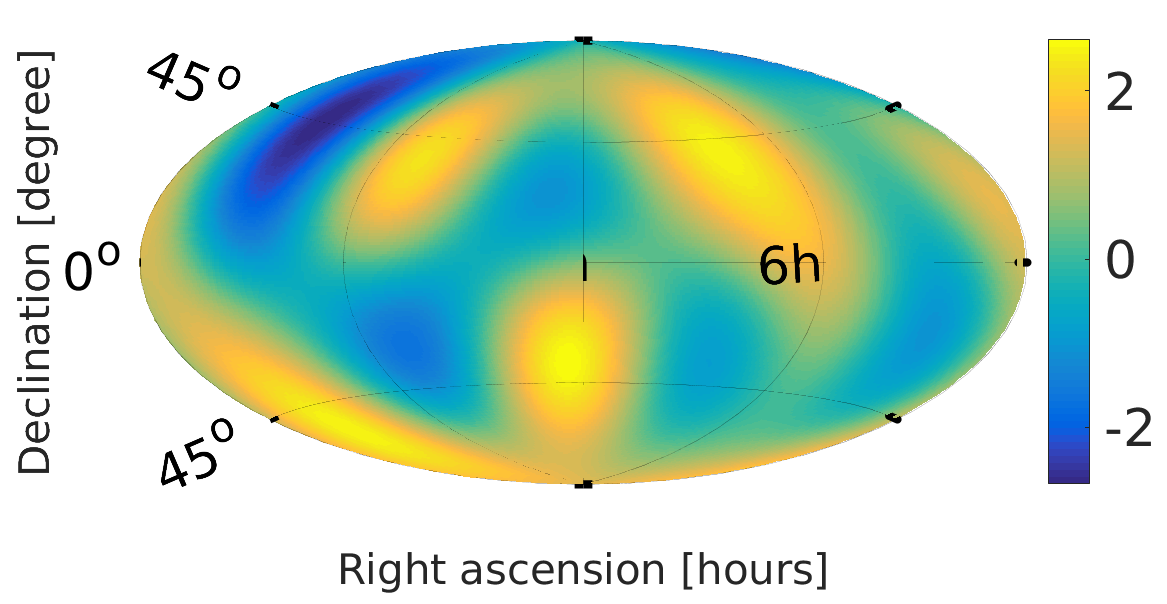} \hspace{.4cm} & 
	\includegraphics[width=2.15in]{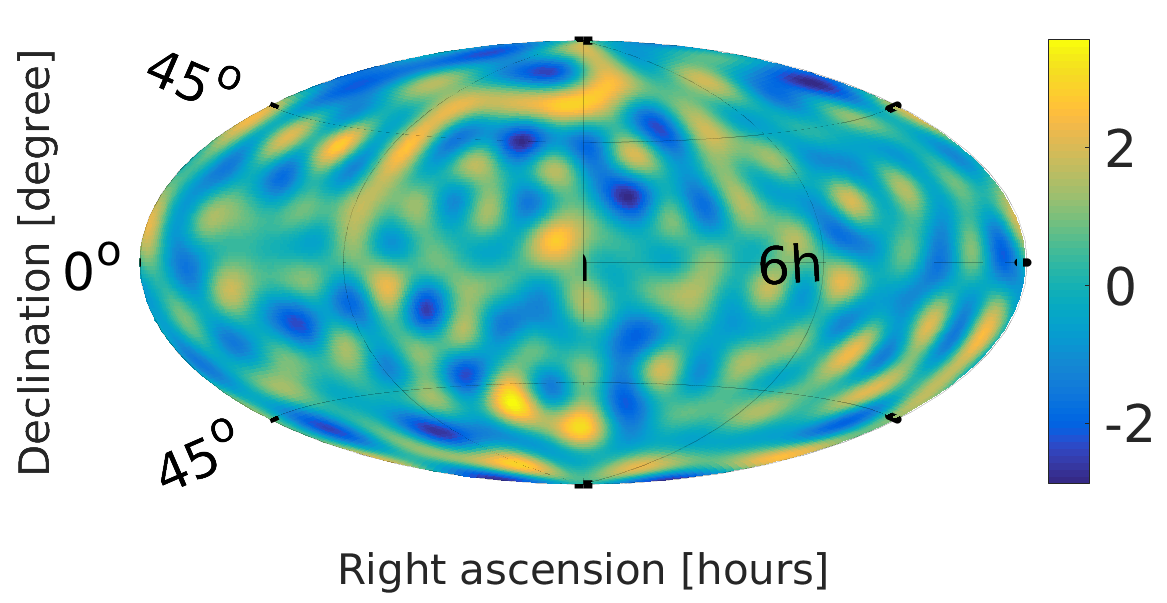} \hspace{.4cm} \\ 
	\includegraphics[width=2.3in]{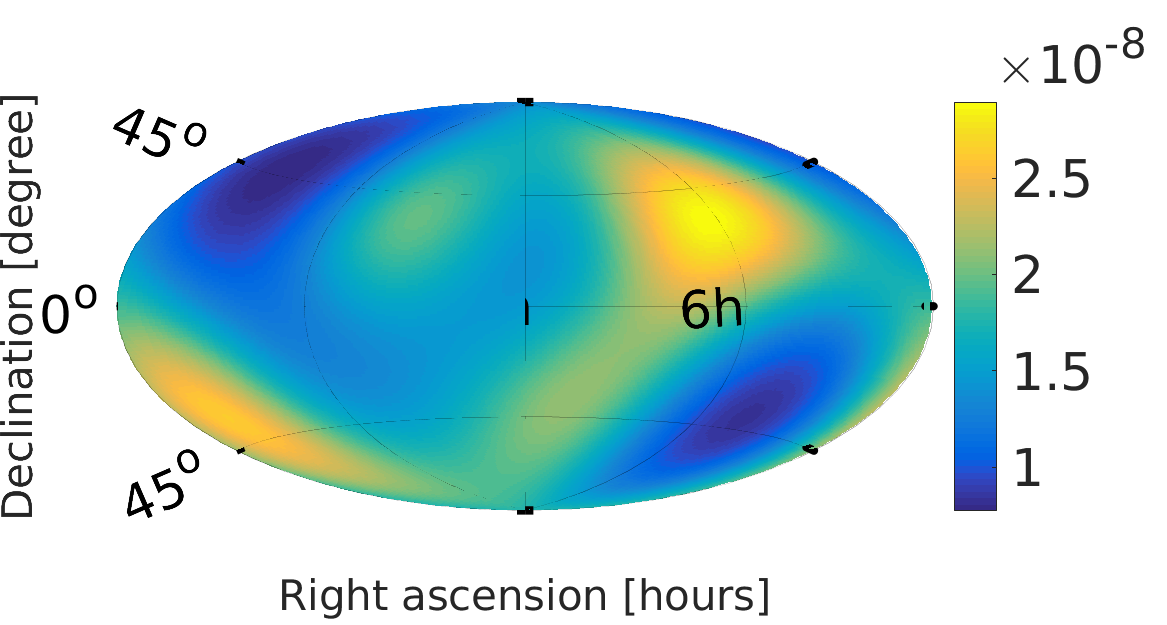} &  
	\includegraphics[width=2.3in]{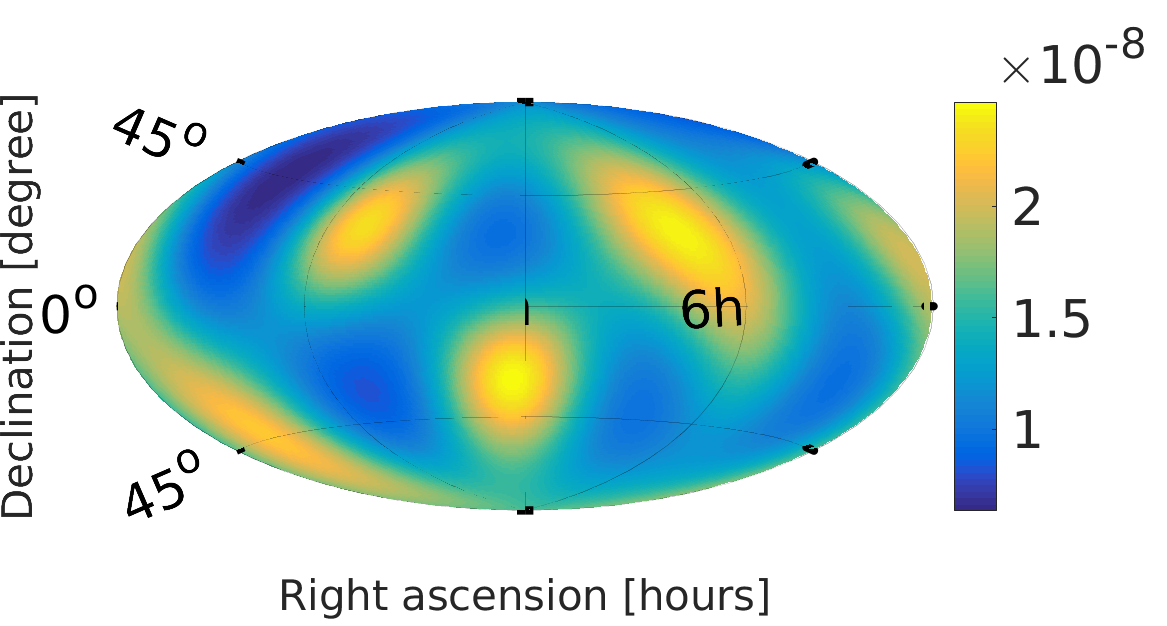} & 
	\includegraphics[width=2.3in]{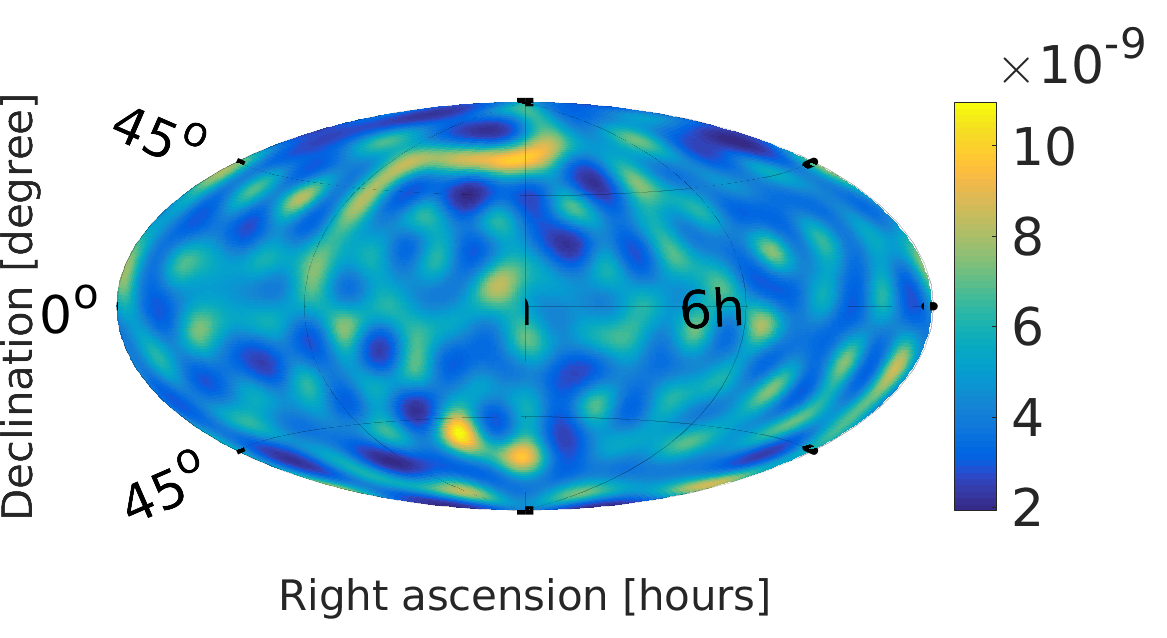} 
  \end{tabular}
  \caption{
All-sky maps reconstructed from a spherical harmonic decomposition. This search is optimized for extended sources, and the plots above show SNR (top) and upper limits at 95\% confidence on the energy density of the SGWB $\Omega_{\alpha}\; [\text{ sr}^{-1}]$ (bottom). Results for three different power-law spectral indices, $\alpha = 0, \, 2/3$ and 3 are shown from left to right. These three different sets of maps have an $l_{\rm max}$ of 3, 4, and 16 respectively.
The $p$-values associated with the maximum \ac{SNR} are (from left to right) $p=9 \%,\,p=31\%,\,p=27\%$ (see Table~\ref{tab:alphas}).}
\label{fig:SPHmaps}  
\end{figure*}

The sky maps for the BBR search are shown in Figure~\ref{fig:RADmaps}, and for
the SHD search in Figure~\ref{fig:SPHmaps}. Converting maps from the spherical
harmonics basis (i.e. $\mu=(lm)$) to the pixel basis is discussed in detail
in~\cite{sph_methods}. Each column indicates a different value of the spectral
index, $\alpha$. The top row shows a map of the signal-to-noise ratio (SNR) for
each sky direction.  The SNR sky maps are consistent with Gaussian noise (see
the $p$-values given in Table~\ref{tab:alphas}).  Consequently, we place upper
limits on the amount of GW power in each pixel using the methods
outlined in~\cite{whelan_calibration_multi_baseline}.  The bottom rows of
Figures~\ref{fig:RADmaps} and ~\ref{fig:SPHmaps} show maps of these upper
limits for the BBR and SHD analyses, respectively. The minimum and maximum 95\%
confidence upper limits across all pixels for both the BBR and SHD searches are
shown in Table~\ref{tab:alphas}. These limits represent a median improvement
across the sky of 2.6-2.7 for the BBR search and 2.8-3 for the SHD search,
depending on the power-law spectral index, $\alpha$.

\subsection{Limits on angular power spectra}\label{ssec:results:shd}
\begin{figure}[h!]
   \includegraphics[width=3.8in]{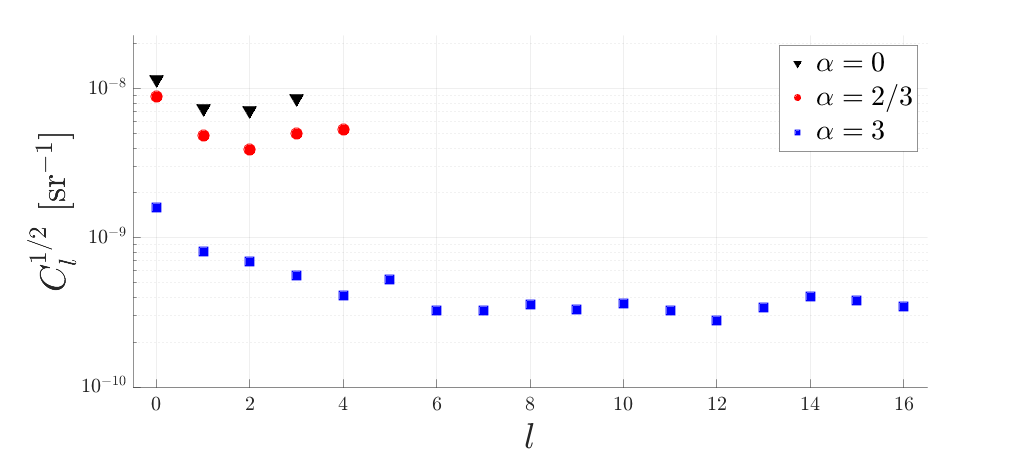}
  \caption{
    Upper limits on $C_l$'s at 95\% confidence for the SHD analyses for $\alpha=0$ (top, black triangles), $\alpha=2/3$ (middle, red circles) and $\alpha=3$ (bottom, blue squares). These represent an improvement in upper limits over O1 of 2.5 -- 3 depending on spectral index, $\alpha$, and $l$.
    \label{fig:Cls}
  }
\end{figure}

We also use the maps from the SHD analysis to set upper limits on the angular
power spectrum components, $C_l$. The upper limits are shown for three spectral
indices in Figure~\ref{fig:Cls}.  The upper limit for $\alpha=2/3$ can be
compared with   theoretical predictions in the literature for the SGWB from
compact binaries~\cite{Jenkins:2018uac,Jenkins:2018kxc,Cusin:2018rsq}.  In
particular, the calculation in Refs.~\cite{Jenkins:2018uac,Jenkins:2018kxc}
gives $C_l^{1/2}\approx3\times10^{-11}~\mathrm{sr}^{-1}$ for $1\le l\le 4$ (the
calculation in Ref.~\cite{Cusin:2018rsq} gives values that are $\sim10\times$ smaller).  Similarly, the
upper limit for $\alpha=0$ can be compared with predictions for the SGWB from
Nambu-Goto cosmic strings in Ref.~\cite{Jenkins:2018lvb}, using the same models
for the string network as in Ref.~\cite{Abbott:2017mem}.  Assuming the
isotropic component of the cosmic string SGWB is consistent with the upper
limits set by LIGO's second observing run~\cite{o2_isotropic_paper}, the dipole ($l=1$) can be as large as
$C_1^{1/2}\approx10^{-10}~ \mathrm{sr}^{-1}$, though the values for higher
multipoles $l>1$ are many orders of magnitude smaller.  These predictions are
therefore consistent with the upper limits obtained here, and present an
important target for future observing runs.

Looking forward towards the prospect of detection, it is important to note that the finite sampling of the galaxy distribution and the compact binary coalescence event rate induces a shot noise in the anisotropies of the astrophysical GW background. As it has been recently shown \cite{jenkins_sakellariadou_shot_noise} this shot noise leads to a scale-invariant bias term in the angular power spectrum $C_l$ and it scales with observing time. Such a bias will dominate over the true cosmological power spectrum, which to be recovered will need either sufficiently long observing times or subtraction of the foreground.


\subsection{Narrowband radiometer results}\label{ssec:results:nbr}
The narrowband radiometer search estimates the strain amplitude, $h_0$, of a
potential source of GWs in three different directions. The maximum SNR across
the frequency band and an estimate of the significance of that SNR for each
direction are shown in Table~\ref{tab:rad_nb}. The uncertainty on the frequency
for the SNR reported in Table~\ref{tab:rad_nb} is a reflection of the original (uncombined) frequency bin width.
The ephemeris for Scorpius X-1 has been updated since the publication
of~\cite{o1_directional}, and so the search presented below assumes a projected
semi-major axis, $a_0$, in the center of the range presented
by~\cite{scox1_ephemerides_2018}.

\begin{figure*}
  \begin{tabular}{ccc}
  \includegraphics[width=2.3in]{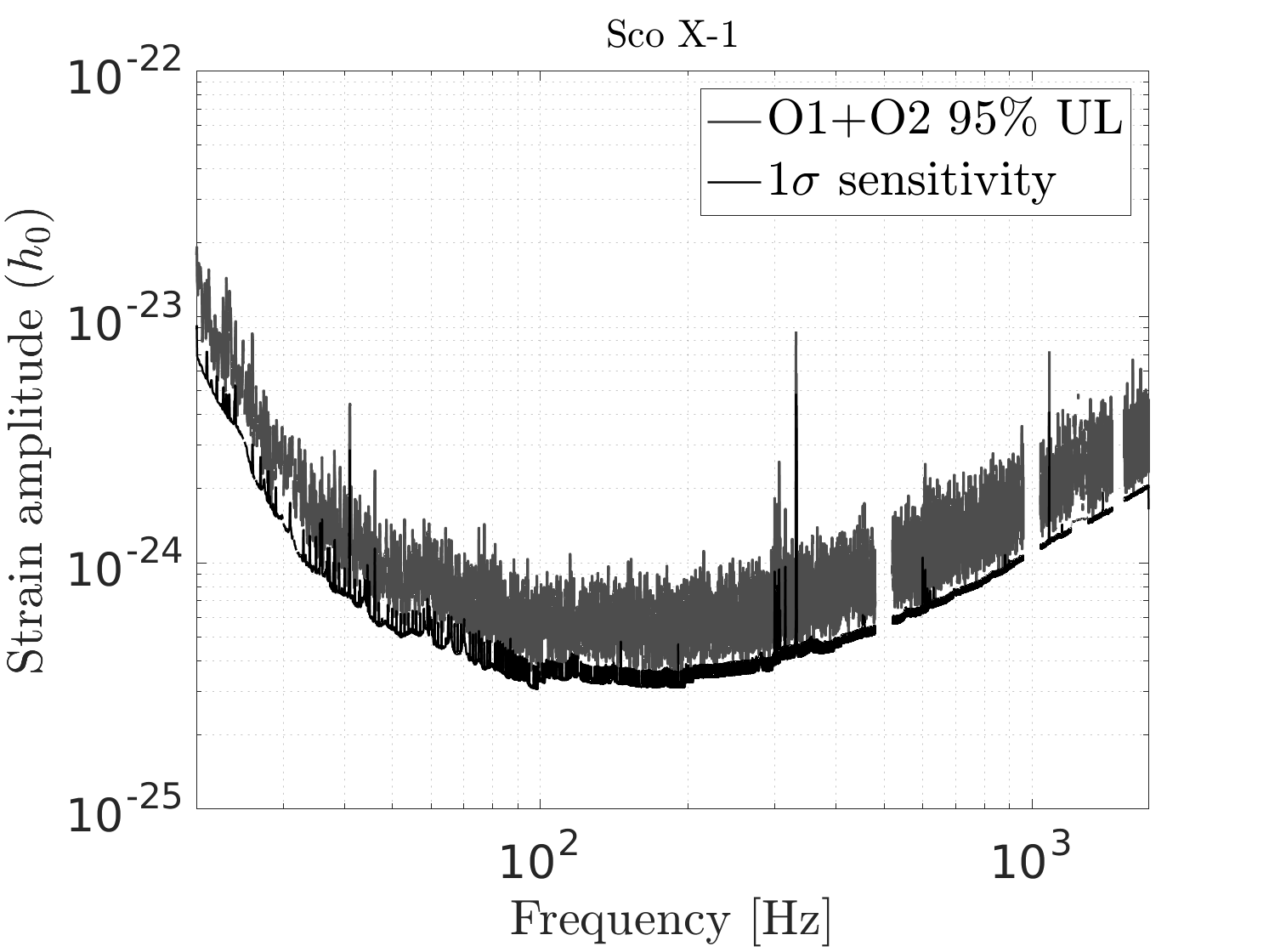} & \hspace{-.4cm}
  \includegraphics[width=2.3in]{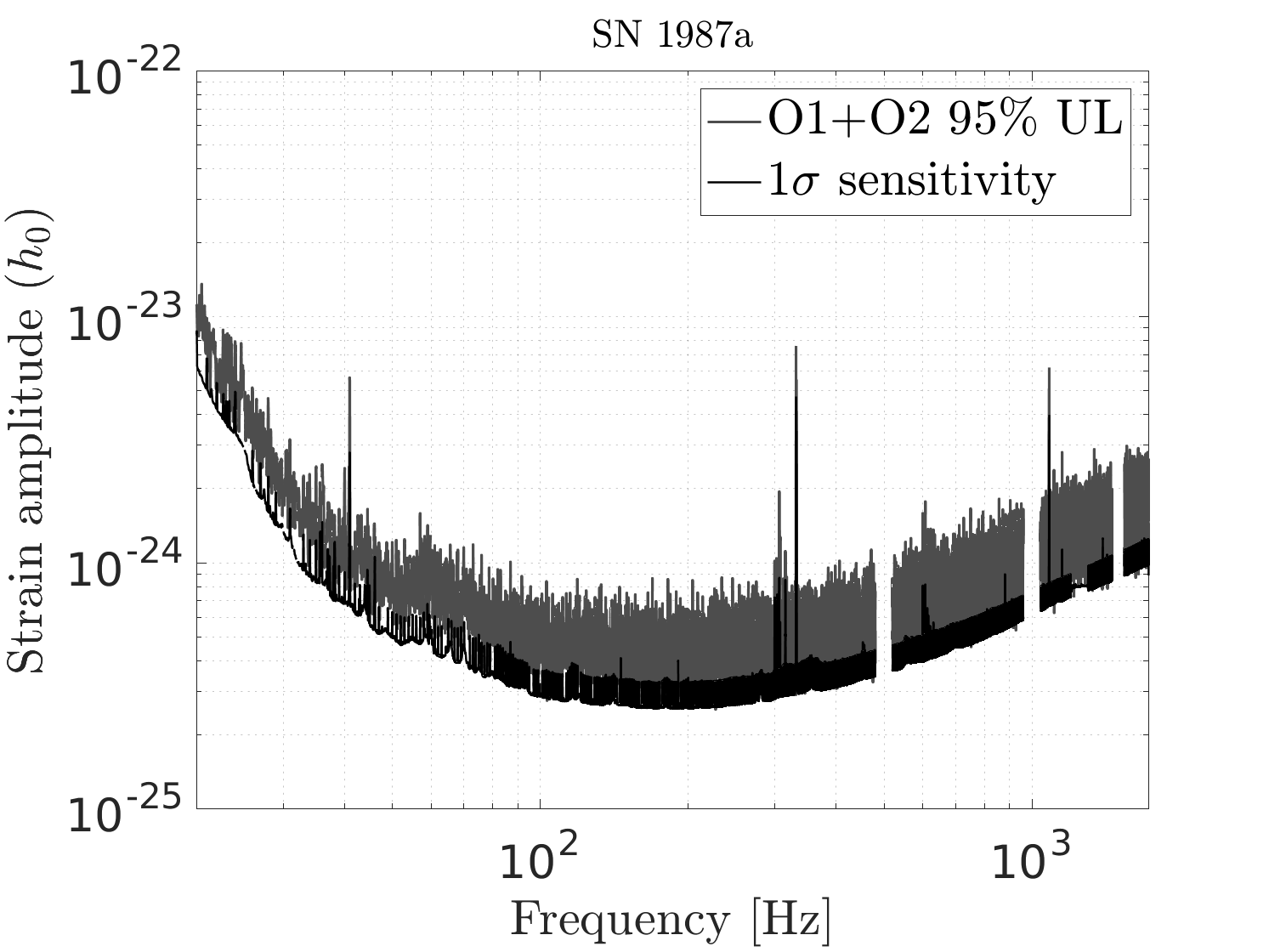} & \hspace{-.4cm}
  \includegraphics[width=2.3in]{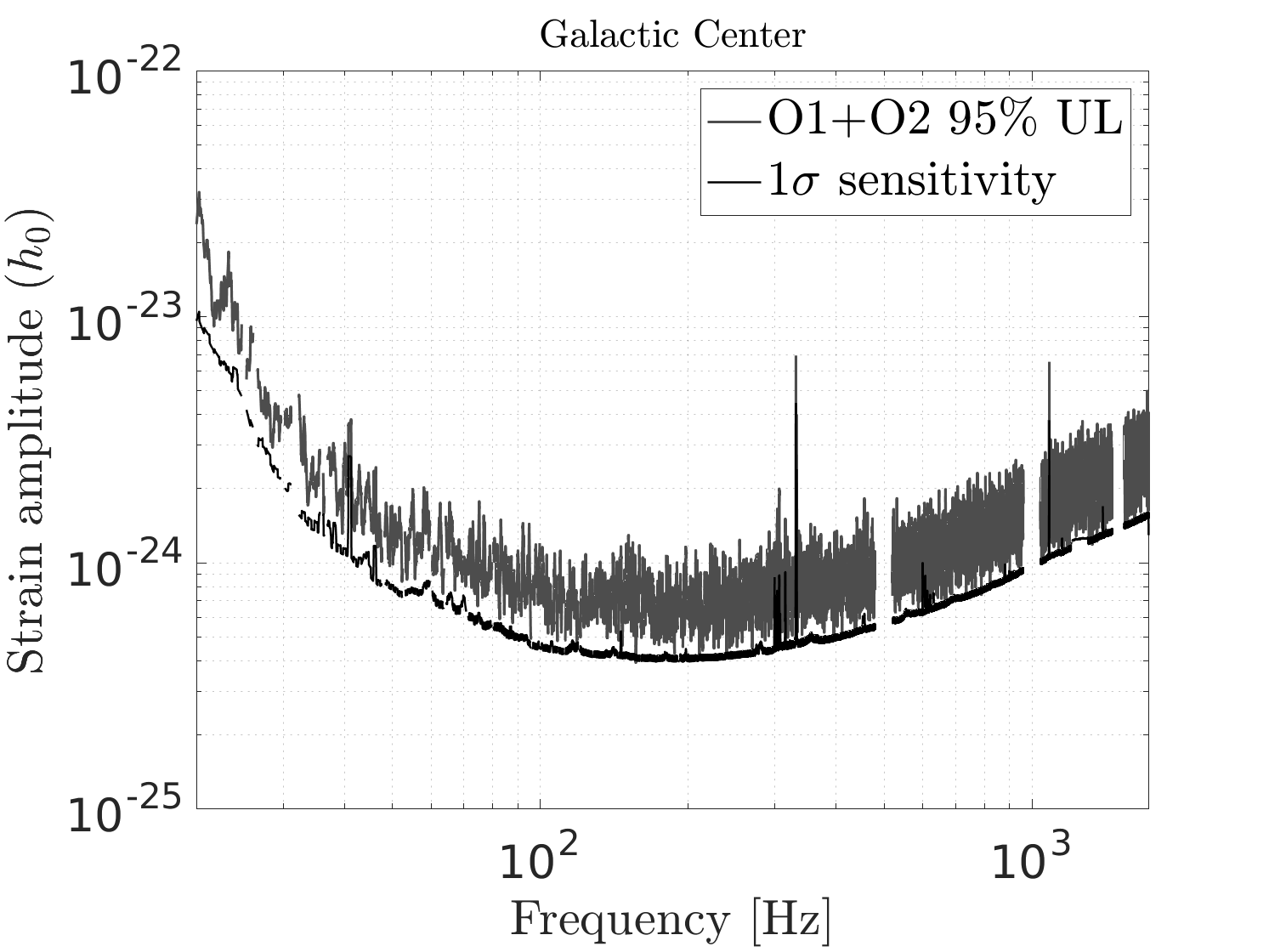} \\
  \end{tabular}
  \caption{
Upper limit spectra using data from O1 and O2 on the dimensionless strain amplitude, $h_0$, at the 95\% level for the narrowband radiometer search are indicated by the gray bands for Sco X-1 (left), SN 1987A (middle) and the Galactic Center (right).  The dark black line indicates the 1$\sigma$ sensitivity of the search in all three directions.
The large spikes are the result of the calibration lines injected into the detector and suspension-wire resonances for various optical elements throughout the instruments.
 } \label{fig:rad_directional}
\end{figure*}

In the direction of Sco X-1 and the Galactic Center, the maximum SNR is
consistent with what one expects from Gaussian noise.  In the direction of SN
1987A, there is a frequency bin with a 1-sided, single-direction $p$-value
1.7\% at \unit[181.8]{Hz}. This $p$-value includes a trials factor for the number of frequency bins in the analysis. Under the assumption that we search over three independent directions,
an extra trials factor would be applied and this $p$-value rises to 5\%. Therefore, we
find no compelling evidence for GWs from the analysis that combines frequency
bins together. We set 95\% upper limits on the strain amplitude of a putative sinusoidal gravitational wave signal, $h_0$, in each individual frequency bin, taking
into account any Doppler modulation in the signal as well as marginalizing over
inclination angle and polarization angle of the source~\cite{o1_directional}. These limits, along
with the 1$\sigma$ sensitivity of the search, are shown in
Figure~\ref{fig:rad_directional}. To avoid reporting our best limits from
downward fluctuations of noise, we take a running median over each \unit[1]{Hz}
frequency band and report the best limit on $h_0$ and the frequency band of
that limit in Table~\ref{tab:rad_nb}.

The best limits on Sco-X1 set in this paper are higher than the best limit set in O1 using a model-based cross-correlation method~\cite{scox1_cross_corr_o1_results}, and are now lower than those set using hidden Markov model tracking~\cite{scox1_viterbi_o1_results}. The torque-balance limit, set by assuming that torque due to accretion is equal to the braking torque due to GW emission, is still around a factor of 5 lower than the limits set in this paper.
The best limits on $h_0$ in the direction of the Galactic Center and SN 1987A are generally higher than previous modeled searches for isolated neutron stars~\cite{SunEA_1987a,gc_2013,Dergachev_o1_loose_gc,Dergachev_loose_all_sky,E_at_H_all_sky_o1} in the frequency bands where those analyses overlap with the one presented here. This search spans a wider frequency band (20-1726~Hz) than any one of those individual analyses. It is important to note that the search presented in this paper is inherently unmodeled, meaning it makes no assumption about the phase evolution of a potential signal past timescales of 192 s.

\begin{table*}
  \begin{tabular}{ccc|ccc|cc}
   \multicolumn{3}{c}{} &\multicolumn{5}{c}{\bf Narrowband Radiometer Results} \\
   \multicolumn{2}{c}{Direction}&& Max SNR & $p$-value (\%) & Frequency (Hz) ($\pm~\unit[0.016]{Hz}$) & Best UL ($\times 10^{-25}$) & Frequency band (Hz) \\
    \hline 
    &Sco X-1    		&& 4.80  & 4.5 & 1602.09   & 4.2 & $183.6 - 184.6$ \\
    &SN 1987A			&& 4.95  & 1.7 &  181.81  & 3.6 & $247.75-248.75$ \\
    &Galactic Center   	&& 3.80  & 98 & 20.28 & 4.7 & $156.8 - 157.8$ \\ \hline
  \end{tabular}
  \caption{Results for the narrowband radiometer search.  We give the maximum SNR, corresponding $p$-value, and the frequency bin of the maxmum SNR for each direction in which we searched. We also give the best 95\% GW strain upper limits achieved, and the corresponding frequency band, for all three sky locations. The best upper limits are taken as the median of the most sensitive 1 Hz band.} \label{tab:rad_nb}
\end{table*}

\section{Outlier at 36.06~{\rm Hz} in the O2 data}
In the process of performing the narrowband radiometer search, a natural
intermediate step of the analysis is to look directly at the 0.03125 Hz
bins for the O2 data, before combining with O1 and before combining over
adjacent bins to account for Doppler modulation. We call these ``sub-bins''.
For this intermediate data
product, the maximum SNRs for the Galactic Center, Sco X-1, and SN 1987A are
4.6, 4.3, and 5.3, respectively.  These first two values correspond to
$p$-values greater than $5\%$, consistent with Gaussian noise.  But for
SN~1987A, the maximum SNR of $5.3$ at 36.0625~Hz has a corresponding $p$-value
of $0.27\%$, or $3\sigma$, which is marginally significant.

Assuming that the maximum SNR is due to a pulsar which 
is spinning down due to GW emission, 
we can relate the observed strain 
$h_0=7.3\times 10^{-25}$ (assuming circular polarization)
at $f=36.06~{\rm Hz}$ to other parameters describing 
the pulsar:
\begin{equation}
    h_0 = \frac{4\pi G}{c^4} \frac{\epsilon I_z f^2}{r},\ \ 
    \dot{f}= -\frac{G}{5\pi c^5} \epsilon^2 I_z (2\pi f)^5.
\end{equation}
We use a fiducial value for the moment of inertia $I_z=10^{39}{\ \rm kg \cdot
m^2}$.  If the source is associated with SN~1987A, then the distance to Earth
is approximately $r=51~{\rm kpc}$~\cite{panagia_sn1987a_review, panagia_1991_distance}, leading to an ellipticity $\epsilon =
3\times 10^{-2}$ and spin down $\dot{f}=-7.7\times 10^{-8}\ {\rm Hz/s}$.  But
this value of the spin down parameter is inconsistent with the fact that the signal is
seen in only one frequency bin.  For the signal to remain in a single frequency
bin, it either needs to have some balancing torque, perhaps from accretion~\cite{Bildsten_1998}, or the signal would need to be at $r \lesssim 1~{\rm kpc}$ (corresponding to 
$\dot{f}=-2.9\times 10^{-11}\ {\rm Hz/s}$). In the latter case, the ellipticity
$\epsilon=5\times 10^{-4}$ is still much larger than that predicted for
typical pulsars. An ellipticity this large is unlikely to be caused by elastic deformations~\cite{elastic_deformation_maximum}, but could in principle be caused by large internal magnetic fields~\cite{gw_emission_magnetic_field_distortion, haskell_model_magnetic_deformation, Melatos_payne_2005}, especially if the protons form a type II
superconductor~\cite{mastrano_melatos}.
It is important to note that an all-sky search for continuous GWs from isolated systems has set limits more stringent than our estimate of $h_0$ for this outlier~\cite{E_at_H_all_sky_o1}.

\begin{figure}
    \centering
    \includegraphics[width=0.49\textwidth]{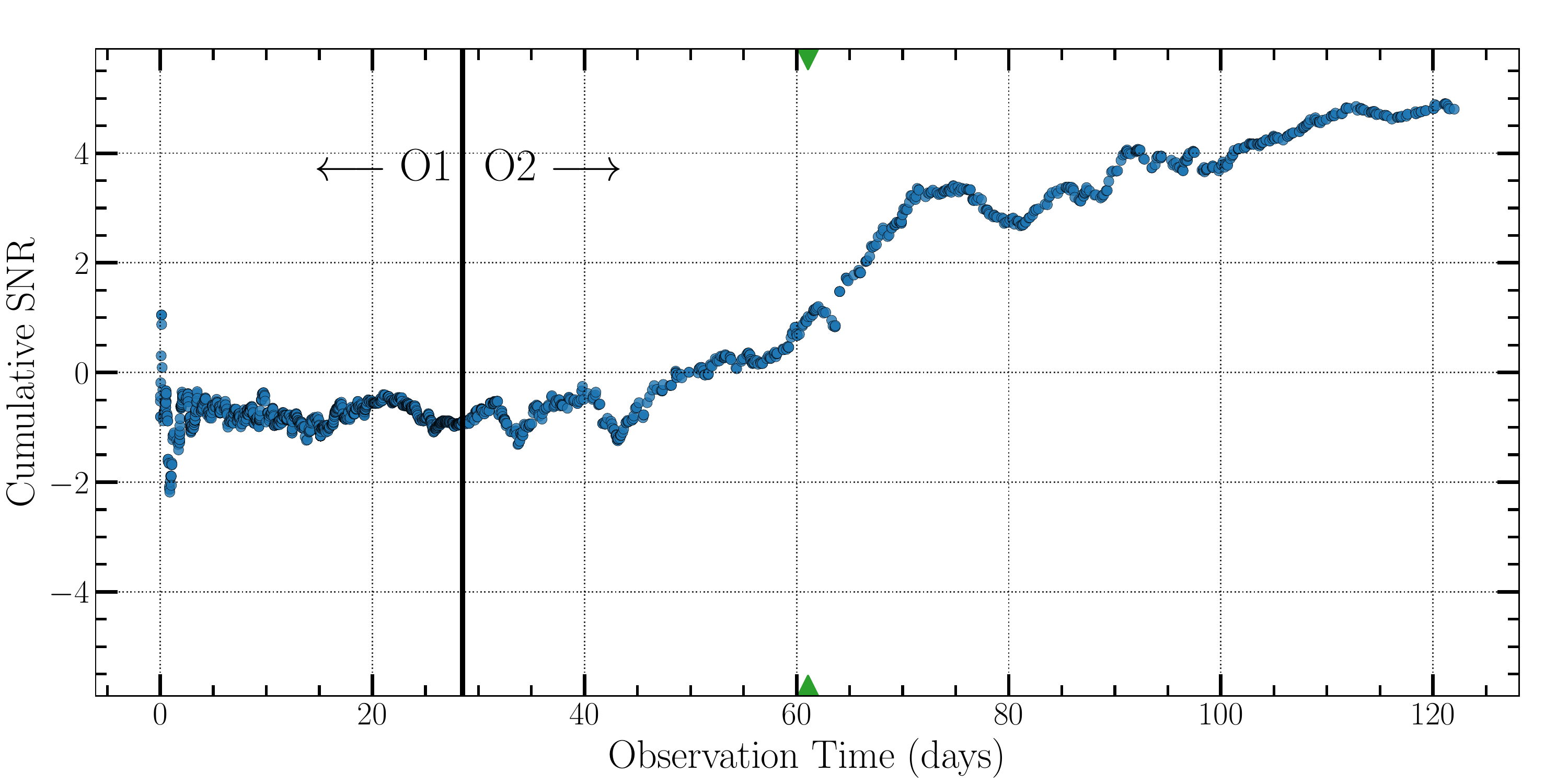}
    \caption{The accumulation of SNR as a function of time for SN 1987a, including both O1 and O2 data. The curve shows the observed cumulative SNR in the 36.06~Hz frequency bin. The green arrows indicate the point where the SNR moves above 1 for the remainder of the available observation time.}
    \label{fig:snr-vs-time}
\end{figure}

Using the techniques described in \cite{o1_o2_lines_paper}, we have not been
able to identify a coherent instrumental witness channel that would explain
this large SNR.  But the fact that the sky direction of the maximum SNR is
close to the equatorial pole is consistent with the behavior of instrumental
noise lines, since the equatorial poles have no sidereal-time modulation.  The
signal appears to turn on during O2, with the SNR exceeding 1 on March 13th, 2017, as shown in Figure
\ref{fig:snr-vs-time}. In addition, the signal does not exhibit any significant short-term
non-stationarity, indicating that this outlier is not generated from a small number of mis-behaved time chunks with large SNRs.
The turn-on feature of the cumulative SNR is not evidence of a real signal, however,
as we have performed simulations of Gaussian noise conditioned on getting a
maximum SNR $\geq 5$, and have found examples where a turn-on like this can be
produced. In addition, upon combining O2 and O1 data together, the SNR of this frequency bin is reduced to 4.7, which corresponds to a $p$-value of 10\%, which is consistent with noise.

\section{Conclusions}\label{sec:conclusions}
We have placed upper limits on the anisotropic SGWB using
three complementary methods. In each case we do not find conclusive evidence
for a GW signal, and so we place upper limits by combining data
from Advanced LIGO's first and second observing runs.  A marginal
outlier at a frequency of 36.06 Hz was seen by the narrowband radiometer search
in O2 in the direction of SN 1987A; however it does not appear in the combined O1+O2 data and is not
consistent with a persistent signal.  We will continue to monitor this
particular frequency bin during the next observing run, taking advantage of the
greater confidence that comes with increased observation periods and more sensitive detectors.

In the future, the anisotropic searches will include data from Advanced Virgo
as well, and can be used to study specific astrophysical models. Additionally,
new algorithms can take advantage of folded data to produce a wider search of
every frequency and sky position~\cite{folding, pystoch,narrowband_folded1, narrowband_folded2}.

\section{Acknowledgements}
The authors gratefully acknowledge the support of the United States
National Science Foundation (NSF) for the construction and operation of the
LIGO Laboratory and Advanced LIGO as well as the Science and Technology Facilities Council (STFC) of the
United Kingdom, the Max-Planck-Society (MPS), and the State of
Niedersachsen/Germany for support of the construction of Advanced LIGO 
and construction and operation of the GEO600 detector. 
Additional support for Advanced LIGO was provided by the Australian Research Council.
The authors gratefully acknowledge the Italian Istituto Nazionale di Fisica Nucleare (INFN),  
the French Centre National de la Recherche Scientifique (CNRS) and
the Foundation for Fundamental Research on Matter supported by the Netherlands Organisation for Scientific Research, 
for the construction and operation of the Virgo detector
and the creation and support  of the EGO consortium. 
The authors also gratefully acknowledge research support from these agencies as well as by 
the Council of Scientific and Industrial Research of India, 
the Department of Science and Technology, India,
the Science \& Engineering Research Board (SERB), India,
the Ministry of Human Resource Development, India,
the Spanish  Agencia Estatal de Investigaci\'on,
the Vicepresid\`encia i Conselleria d'Innovaci\'o, Recerca i Turisme and the Conselleria d'Educaci\'o i Universitat del Govern de les Illes Balears,
the Conselleria d'Educaci\'o, Investigaci\'o, Cultura i Esport de la Generalitat Valenciana,
the National Science Centre of Poland,
the Swiss National Science Foundation (SNSF),
the Russian Foundation for Basic Research, 
the Russian Science Foundation,
the European Commission,
the European Regional Development Funds (ERDF),
the Royal Society, 
the Scottish Funding Council, 
the Scottish Universities Physics Alliance, 
the Hungarian Scientific Research Fund (OTKA),
the Lyon Institute of Origins (LIO),
the Paris \^{I}le-de-France Region, 
the National Research, Development and Innovation Office Hungary (NKFIH), 
the National Research Foundation of Korea,
Industry Canada and the Province of Ontario through the Ministry of Economic Development and Innovation, 
the Natural Science and Engineering Research Council Canada,
the Canadian Institute for Advanced Research,
the Brazilian Ministry of Science, Technology, Innovations, and Communications,
the International Center for Theoretical Physics South American Institute for Fundamental Research (ICTP-SAIFR), 
the Research Grants Council of Hong Kong,
the National Natural Science Foundation of China (NSFC),
the Leverhulme Trust, 
the Research Corporation, 
the Ministry of Science and Technology (MOST), Taiwan
and
the Kavli Foundation.
The authors gratefully acknowledge the support of the NSF, STFC, MPS, INFN, CNRS and the
State of Niedersachsen/Germany for provision of computational resources.


 \newcommand{\noop}[1]{}
%


\iftoggle{endauthorlist}{
  \let\author\myauthor
  \let\affiliation\myaffiliation
  \let\maketitle\mymaketitle
  \title{The LIGO Scientific Collaboration and Virgo Collaboration}
  \pacs{}
  \author{B.~P.~Abbott}
\affiliation{LIGO, California Institute of Technology, Pasadena, CA 91125, USA}
\author{R.~Abbott}
\affiliation{LIGO, California Institute of Technology, Pasadena, CA 91125, USA}
\author{T.~D.~Abbott}
\affiliation{Louisiana State University, Baton Rouge, LA 70803, USA}
\author{S.~Abraham}
\affiliation{Inter-University Centre for Astronomy and Astrophysics, Pune 411007, India}
\author{F.~Acernese}
\affiliation{Universit\`a di Salerno, Fisciano, I-84084 Salerno, Italy}
\affiliation{INFN, Sezione di Napoli, Complesso Universitario di Monte S.Angelo, I-80126 Napoli, Italy}
\author{K.~Ackley}
\affiliation{OzGrav, School of Physics \& Astronomy, Monash University, Clayton 3800, Victoria, Australia}
\author{C.~Adams}
\affiliation{LIGO Livingston Observatory, Livingston, LA 70754, USA}
\author{R.~X.~Adhikari}
\affiliation{LIGO, California Institute of Technology, Pasadena, CA 91125, USA}
\author{V.~B.~Adya}
\affiliation{Max Planck Institute for Gravitational Physics (Albert Einstein Institute), D-30167 Hannover, Germany}
\affiliation{Leibniz Universit\"at Hannover, D-30167 Hannover, Germany}
\author{C.~Affeldt}
\affiliation{Max Planck Institute for Gravitational Physics (Albert Einstein Institute), D-30167 Hannover, Germany}
\affiliation{Leibniz Universit\"at Hannover, D-30167 Hannover, Germany}
% ...

\collaboration{The LIGO Scientific Collaboration and the Virgo Collaboration}
  \newpage
  \maketitle
}

\onecolumngrid
\begin{center}
\pagebreak
{\large\textbf{Supplement To: Directional limits on persistent gravitational waves using data from Advanced LIGO's first two observing runs}} \\[0.5cm]
(The LIGO Scientific Collaboration and Virgo Collaboration)
\end{center}
\vspace{0.55cm}
\twocolumngrid
\setcounter{equation}{0}
\renewcommand{\theequation}{A\arabic{equation}}

In this technical supplement we provide additional formulas to support the main text. This closely follows the discussion given in \cite{o1_directional}. 

Following past analyses, we assume that we can factorize $\Ogw(f,\Theta)$ into frequency and sky-direction-dependent terms
\begin{equation}
    \label{eq:app:omgw_direction_definition_appendix}
    \Ogw (f, \Theta) = \frac{2\pi^2}{3H_0^2}f^3 H(f)\mathcal{P}(\Theta).
\end{equation}
Note that this quantity has units of the dimensionless energy density parameter per steradian. For the radiometer searches it is useful to define a different representation in terms of energy flux,
\begin{equation}
    \label{eq:app:flux_definition_apendix}
    \mathcal{F}(f, \Theta) = \frac{c^3\pi}{4G} f^2 H(f) \mathcal P(\Theta),
\end{equation}
which has units of ${\rm erg\,s^{-1}\,cm^{-2}\,Hz^{-1}\,sr^{-1}}$, where $c$ is the speed of light and $G$ is Newton's gravitational constant.

We use two different representations to estimate the angular power, $\mathcal P(\Theta)$. The {\it{radiometer}} method~\cite{radio_method,Mitra_2008_radio_method2} (in both the broadband (BBR) and narrowband (NBR) applications) is optimized for a small number of resolvable, separated point sources on the sky, and so we estimate the angular power in terms of point sources by decomposing onto delta functions
\begin{equation}
    \label{eq:app:pixel_basis_functions_appendix}
    \mathcal{P}(\Theta) = \mathcal P_{\Theta_0}\delta^2(\Theta, \Theta_0).
\end{equation}
The radiometer method assumes that the sources are well-localized on the sky (that is, to within one pixel), and so it is not well-suited to sources which are spread over a large solid angle.

To characterize diffuse sources of GWs, we use the {\it{spherical harmonic decomposition}} (SHD)~\cite{sph_methods}. We write the angular power in terms of a sum over spherical harmonics, $Y_{lm}(\Theta)$, with amplitude coefficients, $\mathcal{P}_{lm}$
\begin{equation}
    \label{eq:app:shd_decomposition_definition}
    \mathcal{P}(\Theta) = \sum_{l=0}^{\lmax}\sum_{m=-l}^{l} \mathcal{P}_{lm} Y_{lm}(\Theta).
\end{equation}
In principle, $\lmax$ should be infinite. However in practice, we must take a finite value of $\lmax$. The optimal choice for $\lmax$ depends upon the spatial separation and the sensitivity curve of the detectors.
We use the same choice as in the previous analysis \cite{o1_directional}, taking $\lmax=\{3,4,16\}$ for the 3 spectral indices.

By construction, the NBR search looks for signals in a narrow range of frequency bins. On the other hand, for the broadband SHD and BBR searches, we must make an additional assumption about the spectral shape of the source.
We assume that the GW power spectrum takes a power-law form,
\begin{align}
\label{eq:app:hf_powerlaw_definition}
H(f) = \left(\frac{f}{f_{\rm ref}}\right)^{\alpha-3}.
\end{align}
In this case, the power in each direction is characterized by a spectral index, $\alpha$, and the amplitude of the energy density or flux at a given reference frequency, $\fref$. As described in the main text, we choose $f_{\rm ref}=\unit[25]{Hz}$ and search for and set limits on spectral indices of $\alpha=(0,\, 2/3,\, 3)$. 

Given the spectral shape $H(f)$, we define the following quantities which are used to construct the sky maps. For the BBR search it is convenient to consider the flux $F_{\alpha, \Theta_0}$,
\begin{equation}
    F_{\alpha, \Theta_0} \equiv \int d^2\Theta \,\mathcal{F}(f_{\rm ref},\Theta) = \frac{c^3\pi}{4G}\fref^2 \mathcal{P}_{\Theta_0} .
\end{equation}
For the SHD search, we will use the dimensionless energy density per unit sky area
\begin{equation}
    \Omega_\alpha(\Theta) \equiv \Omega_{\rm gw}(f_{\rm ref},\Theta) = \frac{2\pi^2}{3H_0^2}\fref^3 \mathcal{P}(\Theta).
\end{equation}
For more details, see \cite{o1_directional}.

The starting point of the stochastic analysis is the cross-correlation function $C(f;t)$, which is given by
\begin{equation}
    C(f;t) = \frac{2}{T}\tilde s_1^*(f; t) \tilde s_2(f;t),
\end{equation}
where $s_i(f; t)$ is the Fourier transform of length $T$ of the data from detector $i$ at time $t$.
To produce a sky map, we convolve $C(f,t)$ with the generalized
overlap reduction function $\gamma_\mu(f,t)$, which encodes the time delay between the detectors and the  detector response (see \cite{sph_methods} for an explicit definition). We construct the dirty-map $X_\mu$ (see below)
\begin{equation}
    X_\mu = \sum_{f,t} \gamma_{\mu}^*(f,t)\frac{H(f)}{P_1(f;t)P_2(f;t)} C(f;t).
\end{equation}
 Here, $P_i(f; t)$ is the (one-sided) power spectral density of the noise in detector $i$ and $H(f)$ is the chosen spectral model. We use Greek indices $\mu, \nu, \cdots$ to represent angular degrees of freedom. For the SHD search, $\mu,\nu$ run over the spherical harmonic coefficients, e.g., $\mu\equiv (lm)$. For the BBR and NBR searches, $\mu,\nu$ run over individual sky directions (pixels).
 
 The quantity $X_\mu$ is called the ``dirty map'' because it does not faithfully represent the true gravitational-wave power on the sky.
In order to obtain the true power, following \cite{sph_methods} we introduce the Fisher information matrix, $\Gamma_{\mu\nu}$, which encodes the beam pattern of the detector network
\begin{equation}
    \Gamma_{\mu\nu} = \sum_{f,t} \gamma_{\mu}^*(f,t) \frac{H^2(f)}{P_1(f;t)P_2(f;t)}\gamma_\nu (f;t).
\end{equation}

We can get an estimate of the GW power by inverting the Fisher matrix, $\hat{\mathcal{P}}_{\mu\nu} = \Gamma^{-1}_{\mu\nu} X_\nu$. In the case of the BBR and NBR, we ignore correlations between neighboring pixels, and so we don't perform a full matrix inversion, instead taking the inverse of the diagonal elements of the Fisher matrix:

\begin{align}
\hat{\mathcal{P}}_{\Theta} &= (\Gamma_{\Theta\Theta})^{-1} X_{\Theta},\\
\sigma_{\Theta} &= (\Gamma_{\Theta\Theta})^{-1/2}.
\end{align}

In the case of the spherical harmonics, we formally construct an unbiased estimator of the clean map (i.e., the physical map of GW power) using a maximum likelihood estimator \cite{sph_methods}
\begin{equation}
    \hat{\mathcal{P}}_{lm} =\sum_{l'm'} [\Gamma^{-1}_{\rm R}]_{lm,l'm'}X_{l'm'}.
\end{equation}
The Fisher matrix is degenerate because of the existence of blind spots in the detector network, as well as the diffraction limit, \cite{sph_methods}. As a result we need to regularize the Fisher matrix to define an inverse.

Construction of a clean map in the pixel basis, as opposed to the spherical harmonics basis, has been proposed~\cite{Mitra_2008_radio_method2,sph_methods,Renzini_mapping_methods} and implemented on O1 data~\cite{Renzini_o1_mapping_results}. This method also suffers from a poorly-conditioned Fisher information matrix, but given that it avoids a spherical harmonics decomposition entirely, it does not use an initial cut-off in $l_{\rm max}$ as our search does.

\end{document}